\documentclass[journal,twoside,web]{ieeecolor}
\usepackage{jsen}
\usepackage{amsmath,amssymb,amsfonts}
\usepackage{bm} 
\usepackage{algorithmic}
\usepackage{graphicx}
\usepackage{textcomp}
\usepackage{wrapfig}
\usepackage{booktabs}
\usepackage{multirow}
\usepackage[table]{xcolor}   
\usepackage{arydshln}        
\usepackage{subfigure}
\usepackage{mathtools}
\usepackage{gensymb}
\usepackage{commath}
\usepackage{cite}
\usepackage{cleveref}
\usepackage{siunitx}

\def\BibTeX{{\rm B\kern-.05em{\sc i\kern-.025em b}\kern-.08em
		T\kern-.1667em\lower.7ex\hbox{E}\kern-.125emX}}
\definecolor{abstractbg}{rgb}{0.89804,0.94510,0.83137}
\begin{document}
	\title{Temporal Data and Short-Time Averages Improve Multiphase Mass Flow Metering}
	\author{Amanda Nyholm, Yessica Arellano, Jinyu Liu,\\Damian Krakowiak, Pierluigi Salvo Rossi, \IEEEmembership{Senior~Member, IEEE} 
\thanks{This work was partially supported by the Research Council of Norway under the project
PREFERENCE within the PETROMAKS2 framework (project nr. 336355).}
\thanks{A. Nyholm is with the Dept. Electronic Systems, Norwegian University of Science and Technology, Trondheim, Norway (e-mail: amanda.nyholm@ntnu.no).}
\thanks{Y. Arellano is with the Dept. Gas Technology, SINTEF Energy Research, Trondheim, Norway (e-mail: yessica.arellano@sintef.no).}
\thanks{J. Liu and D. Krakowiak are with Dept. Research and Development, KROHNE Ltd., Wellingborough, U.K. (e-mail: j.liu4@krohne.com, d.krakowiak@krohne.com).}
\thanks{P. Salvo Rossi is with the Dept. Electronic Systems, Norwegian University of Science and Technology, Trondheim, Norway, and with the Dept. Gas Technology, SINTEF Energy Research, Trondheim, Norway (e-mail: salvorossi@ieee.org).}
}
    
	\IEEEtitleabstractindextext{%
		\fcolorbox{abstractbg}{abstractbg}{%
			\begin{minipage}{\textwidth}%
				\begin{wrapfigure}[14]{r}{3.05in}%
					\includegraphics[width=3in]{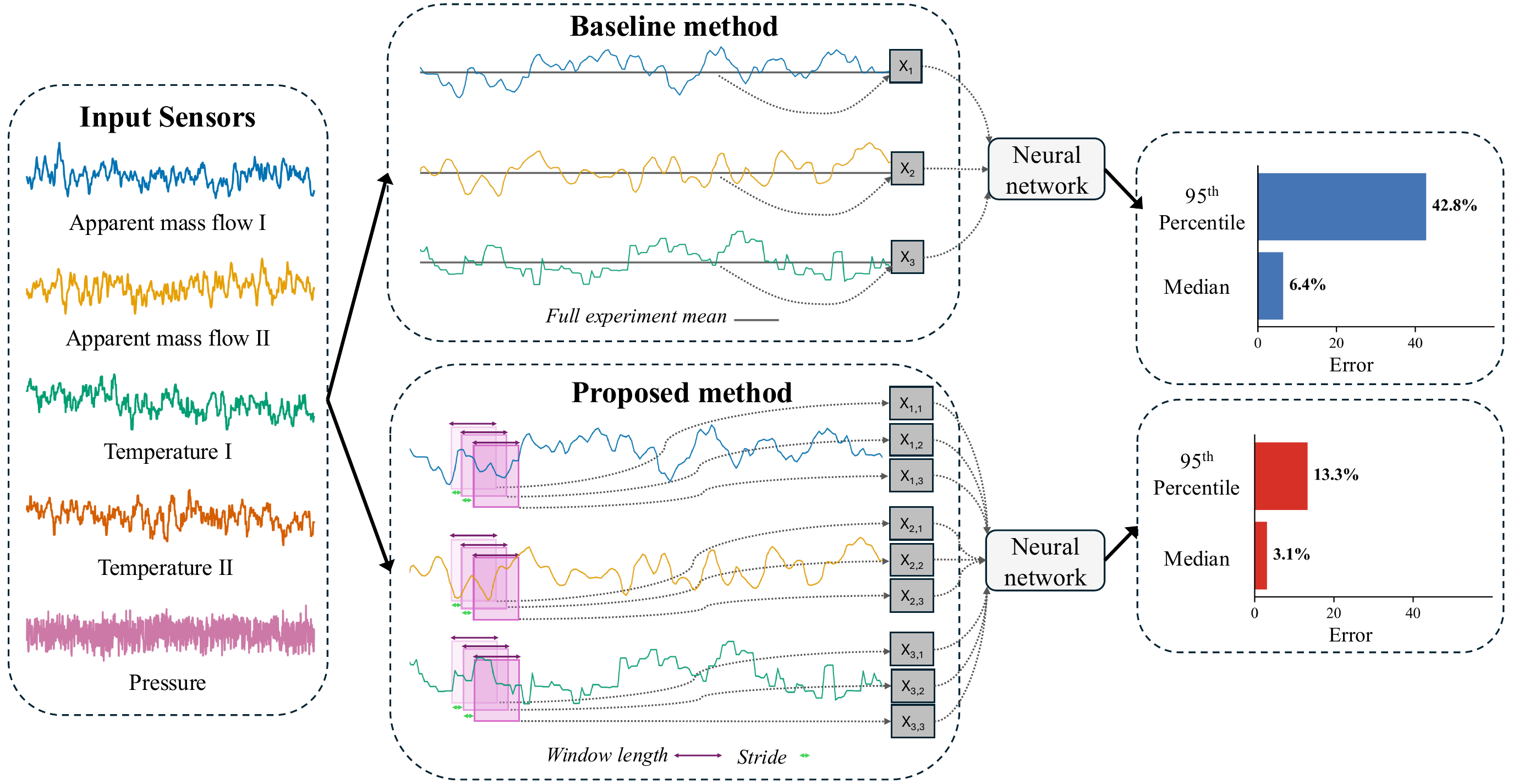}%
				\end{wrapfigure}%
				\begin{abstract}
					Reliable flow measurements are essential in many industries, but current instruments often fail to accurately estimate multiphase flows, which are frequently encountered in real-world operations. Combining machine learning (ML) algorithms with accurate single-phase flowmeters has therefore received extensive research attention in recent years. The Coriolis mass flowmeter is a widely used single-phase meter that provides direct mass flow measurements, which ML models can be trained to correct, thereby reducing measurement errors in multiphase conditions. This paper demonstrates that preserving temporal information significantly improves model performance in such scenarios. We compare a multilayer perceptron, a windowed multilayer perceptron, and a convolutional neural network (CNN) on three-phase air--water--oil flow data from $342$ experiments. Whereas prior work typically compresses each experiment into a single averaged sample, we instead compute short-time averages from within each experiment and train models that preserve temporal information at several downsampling intervals. The CNN performed best at $0.25~\si{\hertz}$ with approximately $95\%$ of relative errors below $13\%$, a normalized root mean squared error of $0.03$, and a mean absolute percentage error of approximately $4.3\%$, clearly outperforming the best single-averaged model and demonstrating that short-time averaging within individual experiments is preferable. Results are consistent across multiple data splits and random seeds, demonstrating robustness. 
				\end{abstract}
				
	\end{minipage}}}
	
	\maketitle
	
\section{Introduction}

\IEEEPARstart{A}{ccurate} measurements of mass flow in mixtures are essential in several industries ranging from \emph{carbon capture \& storage} (CCS) \cite{millsFlowMeasurementChallenges2022} to the food \& beverage \cite{bistaEvaluationValidationInline2019} and oil \& gas \cite{figueiredoUseUltrasonicTechnique2016} sectors. 
These measurements underpin fiscal metering, process control, regulatory compliance, and environmental monitoring, among other applications. 
Mass flow can be estimated using various meter types, and the choice is often application-dependent.

\emph{Venturi meters} (VMs) are differential-pressure meters widely used in chemical and oil \& gas industries to estimate wet gases. They do not provide direct flow rate measurements; instead, they measure pressure differences, from which the volumetric flow rate is inferred using Bernoulli’s principle and then converted to mass flow if the fluid density is known \cite{xuWetGasMetering2011}. 

\emph{Coriolis mass flowmeters} (CMFs) are the only commercial meters that provide direct, highly accurate measurements of both mass flow and fluid density \cite{wangCoriolisFlowmetersReview2014}. Mass flow is estimated by exploiting inertial forces generated as the fluid passes through vibrating tubes, causing a phase shift proportional to the mass flow \cite{arellanoMeasurementTechnologiesPipeline2024}, while the density is determined from the tubes' natural oscillation frequency, which varies with the fluid mass \cite{liuNeuralNetworkCorrect2001}. 
CMFs also provide temperature measurements and additional internal signals, such as tube vibration amplitudes and the oscillation frequency (accessible only to the manufacturer).
    
The high accuracy of CMFs is restricted to single-phase flows, as they assume that the fluid moves homogeneously through the tubes \cite{weinsteinMultiphaseFlowCoriolis2010}. In multiphase conditions, entrained gas or phase separation disturbs the tube oscillations and breaks the proportional relationship with mass flow, leading to large measurement errors, even for low amounts of gas \cite{tombsCoriolisMassFlow2004}. Multiphase mass flowmeters are commercially available, but they are expensive, tend to require extensive maintenance, and some designs rely on radioactive sources \cite{hansenMultiPhaseFlowMetering2019}.

In recent years, the interest in \emph{machine learning} (ML) algorithms for multiphase mass flow estimation has increased significantly \cite{yanApplicationSoftComputing2018,bahramiApplicationArtificialNeural2024}. Two distinct approaches have emerged: (i) using sensor outputs from a single-phase meter, possibly combined with additional sensors to apply corrections; (ii) using measurements from other sensors, such as temperature and pressure, to predict mass flow directly. The latter is often referred to as \emph{virtual flow metering} (VFM). While these approaches are related, they address different underlying problems and are suited for different applications: correction algorithms are, for example, more relevant in industries where CMFs are widely used, whereas VFM is often applied in fields or installations where direct mass flow measurements are impractical. This paper focuses on correction algorithms for CMFs under multiphase conditions. 
    
\subsection{Previous Work}
\subsubsection{Conventional Coriolis Mass Flow Metering Models}

    The first CMF correction algorithm was introduced in 2001 when Liu et al. \cite{liuNeuralNetworkCorrect2001} trained a \emph{multilayer perceptron} (MLP) on an air--water mixture, reducing the \emph{relative error} (RE) from $\pm 20\%$ to $\pm 2\%$ within a limited range of operating conditions. Wang et al. \cite{wangGasliquidTwophaseFlow2016} extended this approach in 2016 with an MLP trained on liquid flow rates between $700$ and $14~500~\si{\kilogram\per\hour}$ and \emph{gas volume fractions} (GVFs)\footnote{GVF represents the proportion of gas in a gas–liquid mixture, where $0\%$ is pure liquid and $100\%$ is pure gas.} up to $30\%$, reducing the RE from $\pm 40\%$ to $\pm 1.5\%$ in vertical pipes and $\pm 2.5\%$ in horizontal pipes. Both implementations used a combination of user-accessible and internal manufacturer signals as inputs.

	Several studies have addressed the correction of CMF estimates in two-phase CO$_2$ mixtures, with the goal of achieving accurate CO$_2$ monitoring in CCS systems. Wang et al. \cite{wangMassFlowMeasurement2017} trained an MLP with only two (user-accessible) features on liquid CO$_2$ flow rates ranging from $300$ to $3050~\si{\kilogram\per\hour}$ and GVFs up to $90\%$, and obtained most REs below $\pm 1.5\%$ (horizontal) and $\pm 2\%$ (vertical). In 2018, they extended this work with a two-stage model on liquid CO$_2$ flow rates from $250$ to $3200~\si{\kilogram\per\hour}$ and GVFs up to $75\%$, where the first stage classified the flow pattern and the second stage predicted the mass flow, yielding REs below $\pm 5\%$ (horizontal) and $\pm 3\%$ (vertical) \cite{wangMassFlowMeasurement2018}. A dynamic ensemble method, which first classified the flow pattern before predicting the mass flow using a least-squares support vector regressor and user-accessible features, achieved REs below $\pm 1\%$ over GVFs ranging from $3.1\%$ to $88.4\%$ and liquid CO$_2$ flow rates between $200$ and $3100~\si{\kilogram\per\hour}$ \cite{sunDynamicEnsembleSelection2018}. Sun et al. \cite{sunNovelHeterogeneousEnsemble2021} used several tree-based algorithms to identify the most important features before training a gradient boosting random forest to predict the mass flow, achieving $70\%$ of REs within $\pm 1.2\%$ across flow rates ranging from $212$ to $3449~\si{\kilogram\per\hour}$, and GVFs between $1.82\%$ and $77.29\%$. A recent study by Wang et al. \cite{wangPerformanceEvaluationCoriolis2024} employed an MLP to correct the mass flow for liquid CO$_2$ injected with gaseous nitrogen and reported that most REs were within $\pm 4\%$. Chowdhury et al. \cite{chowdhuryMassFlowrateMeasurement2024} applied CMF correction algorithms to slurry (sand--water) mixtures using Gaussian process regression, yielding notably low REs below $\pm 0.2\%$ and outperforming both a \emph{support vector machine} (SVM) and an MLP. 

	Henry et al. \cite{henryCoriolisMassFlow2013} fed outputs from a CMF and a water cut (water content in oil) meter into an MLP to predict the mass flow of a three-phase water--oil--nitrogen mixture. The experiments involved high-viscosity oil ($200$ cSt), GVFs between $4.2\%$ and $48\%$, water cuts ranging from $0.2\%$ to $98\%$ and total liquid mass flow rates of $2.5$ to $11.6~\si{\kilogram\per\hour}$. The final model achieved REs below $\pm$ 2.5~\%.
		
	\subsubsection{Sequence Models in Flow Measurements}
    Even under stable conditions, flow measurements exhibit short-term fluctuations, providing a strong motivation for incorporating temporal dependencies into model development. Such models, known as sequence models, are uncommon in CMF corrections but have shown clear benefits in related domains \cite{baoIntegratingMachineLearning2024}, with \emph{long short-term memory} (LSTM) networks proving especially effective in VFM \cite{bikmukhametovFirstPrinciplesMachine2020}.

    Wang et al. \cite{wangMachineLearningMultiphase2020} acknowledged the time-series nature of flow measurements in 2020, and used moving-averaged Venturi outputs to predict volumetric flow rates of a three-phase oil--water--gas mixture using SVM, MLP, and \emph{convolutional neural network} (CNN) models. Building on this, Wang et al. later developed CNN--LSTM and \emph{temporal convolutional network} (TCN) models to predict the volumetric liquid and gas flow rates of the oil--water--gas mixture  \cite{wangMultiphaseFlowrateMeasurement2022}. The same research group further extended their work on the TCN by combining it with a dual-plane electrical capacitance tomography sensor \cite{wangMultiphaseFlowrateMeasurement2023}. Jiang et al. \cite{jiangFlowRateEstimation2024} instead combined an electrical resistance tomography sensor with a Venturi meter to predict the volumetric flow rate of an air--water mixture using a transformer \emph{neural network} (NN) and five-minute-averaged time-series inputs, which outperformed several other sequence models.

    Novel sequence deep-learning models for two-phase gas--liquid mixtures have been proposed in recent years. Bao et al. \cite{baoEnhancingAccuracyGas2024} combined several sensors and developed two sequence models (one based on a \emph{one-dimensional CNN} (1D-CNN) and an LSTM, and another incorporating positional encodings, attention, and a sliding window) for predicting both mass flow and GVF. Gao et al. \cite{gaoMultitaskBasedTemporalChannelwiseCNN2021} proposed a multitask temporal channel-wise CNN for gas void fraction prediction, using four conductance sensors and flow pattern classification as an auxiliary task. Zhang et al. \cite{zhangGasVolumeFraction2025} further advanced this line of work by proposing a dual CNN-transformer mixture NN, augmented with an LSTM, for GVF prediction, which outperformed conventional CNN- and transformer-based models. 
     
    Using sequence models to correct CMF predictions is a logical next step, yet few studies have investigated them. Zhang et al. \cite{zhangDevelopingLongShortTerm2019} demonstrated promising results on a single-phase water mixture using raw tube vibration signals from a CMF as inputs to an LSTM. Sun et al. \cite{sunCoriolisFlowmeterTwoPhase2024} recently compared several sequence models on gas--liquid two-phase data with liquid mass flow rates between $0$ and $1000~\si{\kilogram\per\hour}$ and GVFs of $0\%$ to $30\%$, though without a non-sequence baseline.     
    
\subsection{Contributions}
    Obtaining reliable mass flow rates for individual phases under multiphase conditions is challenging. Single-phase flowmeters are often preferred over more complex multiphase technologies, but their accuracy deteriorates significantly in multiphase flow. CMFs are attractive in multiphase flow conditions because, although their errors are large, they show repeatability. Averaging is often applied to reduce errors, but accuracy remains below industrial requirements. In the system analyzed in this study (described in Section~\ref{sec:data_setup}), raw CMF measurements had a maximum RE of $409\%$, reduced only to $69\%$ after averaging. These limitations motivate further research into ML-based correction algorithms.
    
    While sequence models have been applied in prior work, the sources of their performance gains and their robustness have not been systematically established. 
    This paper addresses these gaps through controlled model comparisons, short-time averaging strategies, and extensive statistical validation on a complex three-phase flow dataset.
    
    This work makes the following contributions:
    \begin{enumerate}
    	\item We demonstrate, through controlled comparisons between two sequence models and a baseline MLP, that preserving temporal information improves CMF correction independently of model capacity.
    	\item We show that short-time averaging within experiments outperforms the current norm of averaging over entire experiments.
    	\item We establish the robustness of the results through multiple evaluation metrics, disjoint data splits, and repeated experiments with multiple random seeds.
    	\item We show that the results hold across a wide range of three-phase flow conditions, including GVFs up to $95\%$, using only user-accessible features to ensure reproducibility.
    \end{enumerate}
    
    This paper is organized into six sections. Section~\ref{sec:data_setup} introduces the dataset and its origin, followed by Section~\ref{sec:preliminaries}, which presents the necessary mathematical preliminaries. Implementation details are given in Section~\ref{sec:implementation}, and the results are discussed in Section~\ref{sec:results}. Finally, Section~\ref{sec:conclusion} summarizes the work and outlines future research directions.

{\em Notation} --- Vectors and matrices are denoted with bold lowercase letters and bold uppercase letters, respectively;
transpose operator is represented with $(\cdot)^{\top}$.

\begin{figure*}[!t]
		\centerline{\includegraphics[width=2\columnwidth]{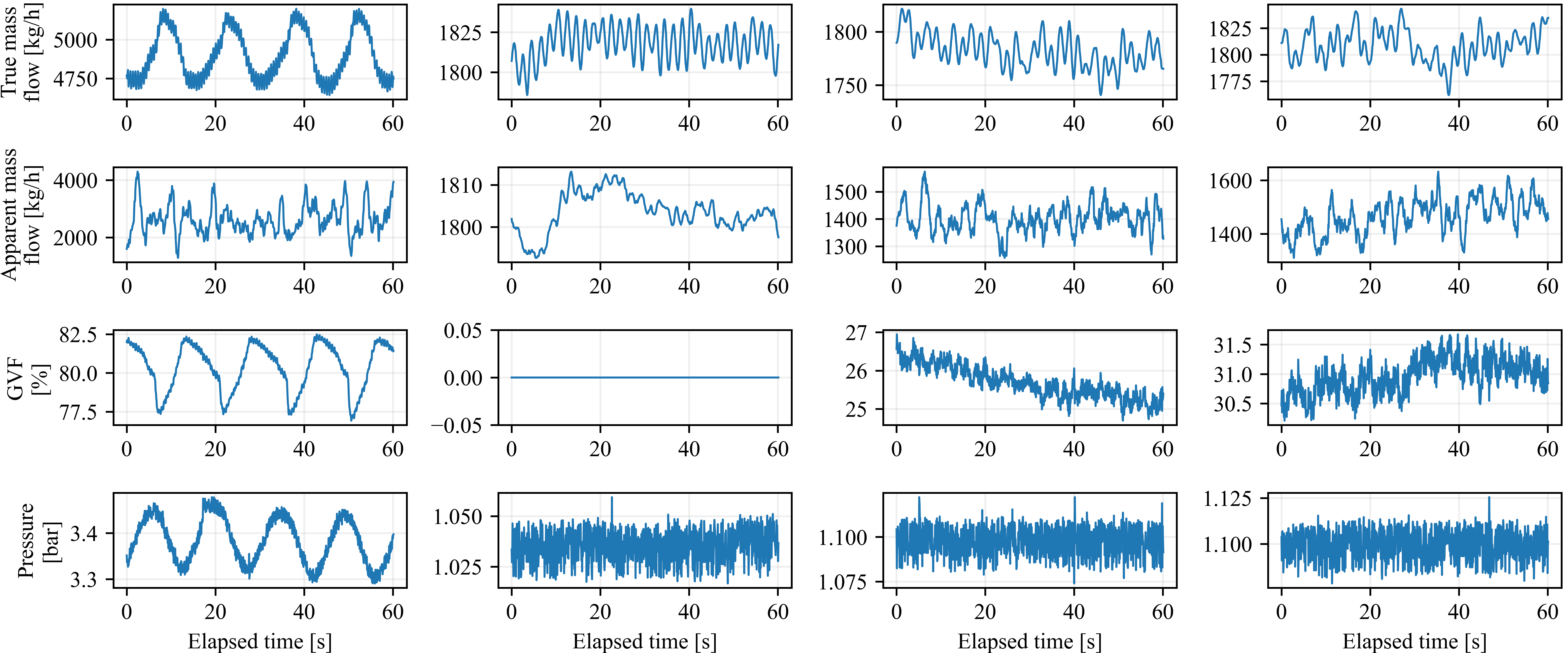}}
		\caption{Representative time series showing (from top to bottom) the true mass flow rate, apparent mass flow rate, GVF, and pressure for four consecutive experiments (columns).}
		\label{fig:overview_of_data}
\end{figure*} 

\section{Experimental Setup and Data}\label{sec:data_setup}
The dataset consists of $294~153$ measurements of a three-phase air--water--oil mixture, collected at $342$ distinct operating points, each hereafter referred to as an experiment. An operating point is defined by a unique combination of water cut, viscosity, oil mass flow rate, total mass flow rate, and GVF. For each baseline setting (water cut, viscosity, oil and total mass flow rates), the GVF was incremented across several consecutive experiments. Each experiment lasted about $60~\si{\second}$ at a fixed GVF and was sampled at $14.3~\si{\hertz}$. Table~\ref{tab:experimental_conditions} summarizes the ranges of the operating variables, and Fig.~\ref{fig:overview_of_data} shows selected measured and reference variables for four consecutive experiments.
    
    \begin{table}[t!]
		\caption{Ranges of operating variables.}
		\setlength{\tabcolsep}{3pt}
		\centering
		\begin{tabular}{llrr}
			\toprule
			\textbf{Variable} & \textbf{Unit} & \textbf{Lower limit}& 
            \textbf{Upper limit} \\
            \midrule
            Pressure & \si{\bar} &  $1.01$ & $4.49$ \\
            Viscosity &\si{\milli\pascal\second} & $7.17  \times 10^{-4}$ & $666$ \\
            Water cut & \% & $0$ & $99.4$\\
            Temperature & \si{\celsius} & $19.2$ & $35.7$ \\
            Oil mass flow rate & \si{\kilogram\per\hour} & $10.6$ & $12,900$\\
            Gas volume fraction & \% & $0$ & $95.5$ \\
            Overall mass flow rate & \si{\kilogram\per\hour} & $930$ &  $14,900$\\
			\bottomrule
		\end{tabular}
		\label{tab:experimental_conditions}
	\end{table}

\begin{figure}[!t]
        \centerline{\includegraphics[width=\columnwidth]{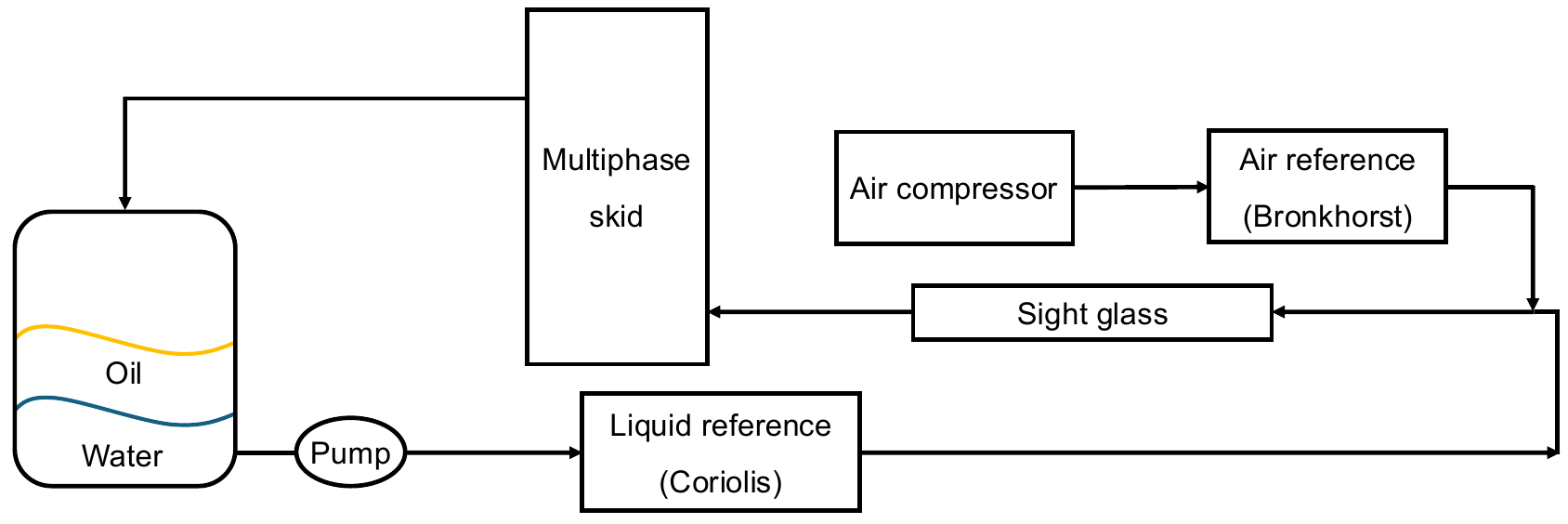}}
        \caption{Illustration of the experimental setup.}
        \label{fig:experimental_setup}
    \end{figure}

\begin{figure}[!t]
        \centerline{\includegraphics[width=\columnwidth]{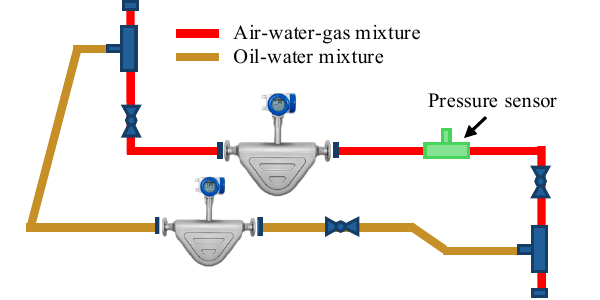}}
        \caption{Illustration of the multiphase skid, rotated 90 degrees clockwise for compactness.}
        \label{fig:skid}
    \end{figure}

    An overview of the experimental setup is shown in Fig.~\ref{fig:experimental_setup}. The multiphase skid was designed to separate the gas from the liquid mixture using gravity before recombining the streams, as illustrated in Fig.~\ref{fig:skid}, in order to obtain representative measurements of the water cut. This type of short-term gas--liquid separation with subsequent remixing is common in multiphase flow measurement systems, particularly within the oil \& gas industry  \cite{thornThreephaseFlowMeasurement2013}. Two CMFs were used to estimate the mass flow, one OPTIMASS 6400 S25 (1 inch) CMF for the air--water--oil mixture, and one OPTIMASS 6400 S08 (1/4 inch) CMF for the oil--water mixture. An additional pressure sensor was placed downstream of the OPTIMASS 6400 S25. The true mass flow was obtained by combining the liquid reference and air reference measurements, whereas the apparent overall mass flow rate was obtained by combining the mass flow outputs of the two CMFs in the skid.
    
	\section{Preliminaries}\label{sec:preliminaries}
    \subsection{Multivariate Time Series}
    A time series is an ordered sequence of measurements, where current values often depend on prior ones (e.g., temperature readings). Modeling time series is done either with statistical methods, such as Kalman filters and ARIMA models, or through ML. It is crucial to prevent data leakage (using future values) when training a model that incorporates temporal information, as this can inflate evaluation metrics on validation and test sets. Creating training/test splits, as well as designing \emph{cross-validation} (CV) schemes, require more care than for models that do not incorporate temporal information.
    
    Furthermore, ML models require specifying the number of past samples to consider, commonly referred to as the window. The window size depends on the problem and the characteristics of the data itself.

    Mathematically, a multivariate time series with $D$ variables and $N$ consecutive measurements is defined as 
    \begin{align}
        \bm{X} &= (\bm{x}[1], \bm{x}[2], \ldots, \bm{x}[N]) \in \mathbb{R}^{D \times N}, \\
        \bm{x}[n] &= (x_1[n], x_2[n], \ldots, x_D[n])^{\top} \in \mathbb{R}^D, \quad n=1, \ldots, N
    \end{align}
    where the $d$th row of $\bm{X}$ corresponds to the values of the $d$th variable and the $n$th column corresponds to the measurements at the $n$th time step.    
    
	\subsection{Multilayer Perceptron Models}
    MLPs are among the simplest ML models and have been referred to under various names---including NNs, artificial NNs, deep NNs, and backpropagation-ANNs---in previous studies \cite{liuNeuralNetworkCorrect2001,wangGasliquidTwophaseFlow2016, wangMassFlowMeasurement2017,wangMachineLearningMultiphase2020,chowdhuryMassFlowrateMeasurement2024}. Despite their simplicity, MLPs have good predictive capabilities, as they combine linear operations with nonlinear activation functions ($\varphi(\cdot)$), allowing them to approximate complex relationships. Their capacity depends on the number of hidden layers ($L$) and nodes per layer ($m$), and they can be written as
    \begin{align}
        y &= \sum_{j=1}^{m_{L}} w^{(L+1)}_{j}\,
              \varphi\!\Bigg(\sum_{i=1}^{m_{L-1}} w^{(L)}_{ji}\,
              \varphi\!\Big(\cdots \Big) + b^{(L)}_j \Bigg) + b^{(L+1)}
    \end{align}
    where $w_{ji}^{(\ell)}$ and $b_j^{(\ell)}$ denote the weights and the biases for the $\ell$th layer, respectively \cite[pp.~33--35, 41--43]{prince2023understanding}. The weights are updated using gradient-based optimizers, such as stochastic gradient descent and Adam. 
    
    \subsubsection{Windowed Multilayer Perceptron}
    The default MLP structure does not inherently account for past samples, limiting its applicability to time-series data. However, using a sliding window, past samples can be provided as inputs, enabling temporal dependencies to be captured by transforming sequences into fixed-size input vectors. 
    We refer to this architecture as MLPw. 
    Formally, we can express the inputs as
    \begin{align}
        \bm{X}_t = \big(\bm{x}_{t-N_w+1},\,\bm{x}_{t-N_w+2},\,\ldots,\,\bm{x}_t\big) \in \mathbb{R}^{D\times N_w} 
    \end{align}
    where $t$ denotes the time index, $N_w$ the window length, and $\bm{x}_t = (x_{1,t},\ldots,x_{D,t})^\top \in \mathbb{R}^D$ the $D$ input features at time step $t$. Note that $\bm{X}_t$ is reshaped into a one-dimensional vector before being passed to the MLPw.
    
    \subsection{Convolutional Neural Networks}
    In contrast to MLPs, CNNs capture relationships between neighboring inputs, enabling better identification of local structures, which explains their extensive use in image processing \cite[p.~61]{prince2023understanding}. They use a collection of convolutional layers to identify the structures, followed by pooling layers to reduce dimensionality and computation, and finally fully connected layers to produce the outputs \cite{oshea2015introductionconvolutionalneuralnetworks}.

    In addition to their traditional use, CNNs can also be applied to time-series data using a 1D-CNN, where the convolution is performed along the temporal domain \cite{baoIntegratingMachineLearning2024}. Compared to other sequence ML models such as LSTMs and transformers, 1D-CNNs are relatively simple. 
    
    The output of one \emph{1D convolutional layer} (Conv1D) at position $i$ is
    \begin{align}
        h_i = \varphi\!\left(\beta + \sum_{k=1}^{K} \omega_k \, x_{i+k-1}\right)
    \end{align}
    where $x$ is the input, $\omega_k$ are the kernel weights of length $K$, $\beta$ is the bias, and $\varphi(\cdot)$ is the nonlinear activation function \cite{prince2023understanding}.
    
	\subsection{Evaluation Metrics}
    
    REs are commonly used to evaluate model performance in mass flow estimation because measurement magnitudes vary widely. 
    For the $n$th measurement, let $y_n$ be the true value, $\hat{y}_n$ the predicted value, and $e_n = \hat{y}_n - y_n$ the corresponding error. 
    The RE is then computed on the original scale as
    \begin{align}
    	\text{RE}_n = \frac{e_n}{y_n} \times 100~\%.
    \end{align}
    They are the main evaluation metric in this study, with the 95th percentile taken as the primary basis for model comparison, being often the most crucial in industrial applications.
        
    The \emph{average normalized root mean squared error} (avgNRMSE), \emph{range normalized root mean squared error} (rNRMSE), \emph{mean absolute error} (MAE), \emph{median absolute error} (MedAE), \emph{mean absolute percentage error} (MAPE), \emph{median absolute percentage error} (MedAPE), and $R^2$ are also used to obtain a comprehensive evaluation of the models:
    \begin{align}
        &\text{RMSE}     = \sqrt{\tfrac{1}{N}\sum_{n=1}^N e_n^{\,2}}, 
        & R^2      &= 1 - \tfrac{\sum_{n=1}^N e_n^{\,2}}{\sum_{n=1}^N (y_n-\bar y)^{2}}, \nonumber\\
        &\text{rNRMSE}   = \tfrac{\text{RMSE}}{\max(\bm{y})-\min(\bm{y})}, 
        & \text{MAE}    &= \tfrac{1}{N}\sum_{n=1}^N |e_n|, \nonumber\\
        &\text{avgNRMSE} = \tfrac{\text{RMSE}}{\bar y}, 
        & \text{MedAE}  &= \operatorname{med}\!\big(|\bm{e}|\big), \nonumber\\
        &\text{MAPE}     = \tfrac{100\%}{N}\sum_{n=1}^N \big|\tfrac{e_n}{y_n}\big|, 
        & \text{MedAPE} &= \operatorname{med}\!\big(\tfrac{|\bm{e}|}{\bm{y}}\big)\times 100\% \nonumber
    \end{align}
    where $\bm{e} = (e_1, e_2,\ldots, e_N)$,  $\bm{y} = (y_1, y_2,\ldots, y_N)$, and $\bar y$ is the mean of $y$.
    
    The NRMSE is preferred to the standard MSE as it enables comparisons across data sets with different scales and can be normalized in different ways. The rNRMSE indicates how the average RMSE compares to the whole data range, whereas the avgNRMSE relates it to the mean of the data. Both metrics are included since the original data exhibit large variability, and we require the model performing well in both scenarios.
    
\section{Implementation}\label{sec:implementation}
	An initial aliasing assessment showed that all features were at risk of aliasing, particularly the mass flow readings when the data were downsampled to $1~\si{\hertz}$ or lower. Consequently, a low-pass FIR filter with a length of $129$ samples and a cutoff frequency set to $80\%$ of the Nyquist frequency was used to reduce aliasing prior to downsampling.  
    
    Each experiment was assigned a group number to easily identify different experimental conditions, and the groups were defined by the chronological order of the experiments. In other words, the first experiment was assigned to group~1, the second to group~2, and so forth, creating an ``even" and ``odd" split. Nine groups were excluded because they contained invalid or too few measurements. The final even split consisted of $148~780$ measurements and $173$ groups, whereas the odd split had $145~340$ measurements and $169$ groups. 
    
    Training and test splits were based on the group number, as the stepwise nature of the experimental setup ensured that both contained a variety of conditions. To ensure representative and sound results, models were trained and evaluated on both splits. From the test split, one-third was set aside as a hold-out validation set to monitor early stopping when training the final model, and was not used in the final test evaluation. Both the validation and test sets thus contained only flow conditions unseen during training and fine-tuning, ensuring generalizability. Although training on $\sim50\%$ of the data is unusual, it was deemed necessary to evaluate the model across a wide range of conditions. Splits were normalized (min-max) using the scaling parameters of the training set. 

    To ensure consistency and robustness of the results, all models were trained on $30$ different seeds using the same splits and hyperparameters. Five features were used as inputs for the models; the temperature and apparent mass flow outputs from both CMFs in the skid, and the readings from the pressure sensor. Models were trained at multiple downsampling intervals ($250~\si{\milli\second}$, $500~\si{\milli\second}$, $1~\si{\second}$, $2~\si{\second}$, $3~\si{\second}$, $4~\si{\second}$, $5~\si{\second}$, $6~\si{\second}$), and compared with models trained on the original data. Group-aware five-fold CV with a rolling window was used to tune the hyperparameters on the training set and avoid data leakage. 

    Models were trained using the MSE loss function, the \emph{rectified linear unit} (ReLU) activation function, and the AdamW optimizer, for up to $60$ epochs, with early stopping (patience = $10$). Different network structures were explored; however, the final MLP models used one hidden layer, with $8$ hidden nodes for the baseline and $16$ hidden nodes for the MLPw. The final network structure of the CNN is shown in Table~\ref{tab:CNN_architecture}. Initial window lengths were selected to capture approximately one dominant oscillation in the apparent mass flow signal from the liquid reference meter while retaining sufficient samples per experiment for training. They were later fine-tuned. Samples discarded during window initialization in the sequence models were likewise removed for the MLP, so that all models were trained and evaluated on identical data.

    \begin{table}[t!]
        \caption{Final convolutional neural network architecture.}
        \setlength{\tabcolsep}{3pt} 
        \centering
        \begin{tabular}{ll}
            \toprule
            \textbf{Layer} & \textbf{Configuration} \\
            \midrule
            Input                & $B \times N_w \times 5$ (five features, $N_w$ timesteps) \\
            Conv1D + ReLU        & $5 \to 16$, kernel size = 3, padding = 1 \\
            Conv1D + ReLU        & $16 \to 8$, kernel size = 3, padding = 1 \\
            Global AvgPool1D     & Output: $B \times 8$ \\
            Fully connected + ReLU & $8 \to 16$ \\
            Fully connected (output) & $16 \to 1$ \\
            \bottomrule
              \multicolumn{2}{p{0.9\columnwidth}}{
              	$B$: batch size. $N_w$: window length. Conv1D: one-dimensional convolution. ReLU: rectified linear unit activation function. AvgPool1D: one-dimensional average pooling.} 
        \end{tabular}
        \label{tab:CNN_architecture}
    \end{table}

    Implementation was done in PyTorch on a MacBook Air with an Apple M2 chip ($8$-core CPU) and $16$~GB of memory.
    
\section{Results and Discussion}\label{sec:results}
	Downsampling the data yielded consistently better models across all metrics than training on the original sampling rate. Decreasing the sampling frequency improved the model performance, peaking at $4~\si{\second}$, although the $3~\si{\second}$ and $5~\si{\second}$ intervals yielded similar results. Only the results from models sampled at $4~\si{\second}$ are discussed henceforth---denoted as ``4s''---as the other rates do not provide additional insights.  The models trained on data averaged across entire experiments are denoted ``60s''.    
		
	The fine-tuned hyperparameters are listed in Table~\ref{tab:hyper_parameters}. Sufficient variations were considered to confirm the robustness of the results, although further gains are likely with more extensive hyperparameter tuning, particularly of window lengths and network architectures.
	
	\begin{table}[t!]
		\caption{Hyperparameters for the final models.}
		\setlength{\tabcolsep}{5pt}
		\centering
		\begin{tabular}{llcccccc}
			\toprule
			& \multirow{2}{*}{\textbf{Model}} & \multirow{2}{*}{\shortstack{\textbf{Window} \\ \textbf{length}}}& \multirow{2}{*}{\shortstack{\textbf{Batch} \\ \textbf{size}}} & \multicolumn{2}{c}{\textbf{Learning rate}} & \multicolumn{2}{c}{\textbf{Weight decay }}\\
			\cmidrule(lr){5-6}
			\cmidrule(lr){7-8}
			& & & & Even & Odd & Even & Odd \\
			\midrule
			\multirow{3}{*}{4s}& MLP       &   -  &   2  & $10^{-3}$ & $10^{-3}$ & $10^{-4}$  & $10^{-4}$   \\
			& MLPw   &  5  &  2  & $10^{-4}$ & $10^{-4}$ &$10^{-4}$  & $10^{-5}$  \\
			&CNN       &  5  &  2  &  $10^{-4}$  & $10^{-4}$  & $10^{-4}$ & $10^{-3}$   \\
			
			\midrule
			\multirow{3}{*}{60s}& MLP       &   -  &   2  & $10^{-3}$ & $10^{-3}$ & $10^{-3}$  & $10^{-5}$   \\
			& MLPw   &  2  &  2  & $10^{-3}$ & $10^{-3}$ &$10^{-3}$  & $10^{-3}$  \\
			&CNN       &  2  &  2  &  $10^{-3}$  & $10^{-3}$  & $10^{-5}$ & $10^{-5}$   \\
			\bottomrule
			
		\end{tabular}
		\label{tab:hyper_parameters}
	\end{table}
	
\subsection{Model Performance}
    \subsubsection{Relative Errors}
    Table~\ref{tab:relative_errors} shows a collection of metrics related to the REs, computed over all 30 seeds rather than per-seed averages. For example, the ``Max'' value corresponds to the maximum error observed across all seeds. 
    The CNN outperforms both MLPs in terms of overall performance and outlier robustness, regardless of the split, while the MLPw in turn significantly outperforms the standard MLP. These results demonstrate that sequence models perform better than non-sequence ones when correcting CMF measurements. This is particularly evident in the comparison between the MLP and MLPw, which share the same network architecture; the only distinction is that MLPw has access to short-term temporal dependencies, resulting in markedly improved performance.
	
	\begin{table}[t!]
		\setlength{\tabcolsep}{3pt}
		\caption{Relative error metrics (maxima, percentiles, and coverage within 10\% and 5\%)  computed from all errors over 30 seeds (no per-seed averaging). Best results in bold; values in \%.}
		\centering
		\label{tab:relative_errors}
		\begin{tabular}{lll cc>{\columncolor{yellow!20}}ccccc}
            \toprule
            \textbf{Split} & \multicolumn{2}{c}{\textbf{Model}}  
            & \textbf{Max} & \textbf{p99} & \textbf{p95} & \textbf{p50} 
            & $\leq$ \textbf{10\%} & $\leq$ \textbf{5\%} \\
            \midrule
            \multirow{9}{*}{Even}
            & \multirow{2}{*}{MLP}      & 4s  & 146.2 & 39.4 & 24.6 & 4.93 & 73.9 & 50.5 \\
            &  & 60s & 86.4  & 59.7 & 42.8 & 6.42 & 62.3 & 40.4 \\
            \addlinespace
            & \multirow{2}{*}{MLPw}   & 4s  & 53.5  & 26.1 & 16.2 & 3.58 & 86.3 & 63.8 \\
            & & 60s & 205 & 62.4 & 25.1 & 5.31 & 70.3 & 48.1 \\
            \addlinespace
            & \multirow{2}{*}{CNN}    & 4s  & \textbf{44.4} & \textbf{22.1} & \textbf{13.3} & \textbf{3.11} & \textbf{90.8} & \textbf{69.6} \\
            &  & 60s & 158 & 41.8 & 16.1 & 5.27 & 80.4 & 47.7 \\
            \addlinespace
            & \multirow{2}{*}{Original} & 4s  & 168 & 84.5 & 47.1 & 8.57 & 54.5 & 38.1 \\
            &  & 60s & 69.4  & 57.1 & 48.5 & 9.50 & 52.0 & 31.4 \\
            \midrule
            \multirow{9}{*}{Odd}
            & \multirow{2}{*}{MLP}      & 4s  & 146 & 39.4 & 24.5 & 4.93 & 73.9 & 50.5 \\
            &  & 60s & 100 & 55.0 & 25.0 & 5.50 & 68.5 & 47.3 \\
            \addlinespace
            & \multirow{2}{*}{MLPw}   & 4s  & 63.6  & 34.6 & 17.0 & 3.58 & 85.0 & 64.2 \\
            &  & 60s & 65.6  & 45.7 & 28.2 & 4.93 & 69.9 & 50.7 \\
            \addlinespace
            & \multirow{2}{*}{CNN}  & 4s  & 55.8  & \textbf{24.0} & \textbf{12.9} & \textbf{2.87} & \textbf{90.9} & \textbf{71.9} \\
            &  & 60s & \textbf{44.5} & 25.7 & 17.8 & 5.17 & 75.1 & 48.2 \\
            \addlinespace
            & \multirow{2}{*}{Original} & 4s  & 239 & 78.9 & 51.8 & 6.83 & 58.3 & 43.1 \\
            &  & 60s & 50.6  & 46.6 & 41.9 & 10.0 & 49.5 & 30.5 \\
            \bottomrule
        \end{tabular}
	\end{table}
	
	Using shorter averages from each experiment yielded lower REs compared to taking one average across an entire experiment. Table~\ref{tab:relative_errors} demonstrates that the 4s models outperform their 60s counterparts, especially for the sequence models. Only the maximum error of the odd 60s CNN is lower than the 4s CNN. However, when compared to the original CMF outputs (Table~\ref{tab:relative_errors}, Max column), the maximum error was reduced by $6$ and $183$ percentage points for the 60s and 4s CNN models, respectively, i.e. the 4s CNN still performed well (especially since occasional outliers are generally acceptable).
	
\begin{figure*}[!t]
    \centerline{\includegraphics[width=2\columnwidth]{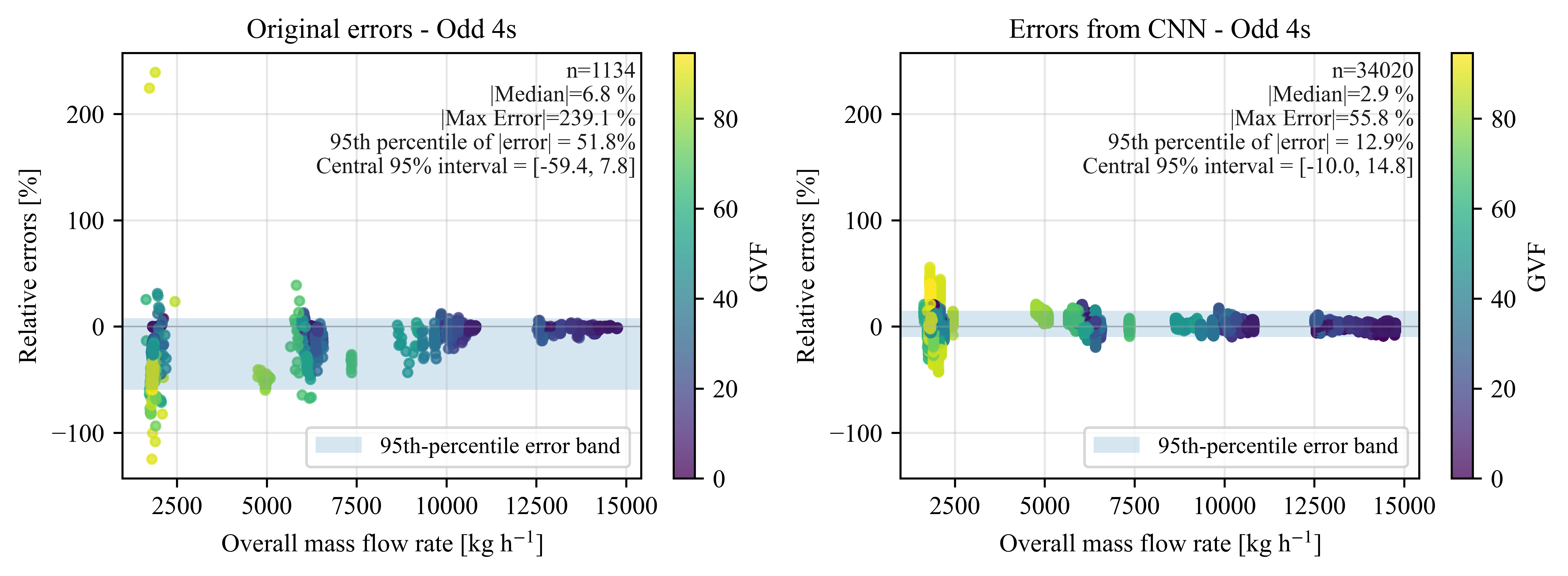}}
    \caption{Relative errors versus overall mass flow rate for the best odd model, compared with the original errors on the same test points.}
    \label{fig:sub_odd}
    \end{figure*}

	The best model (4s CNN) shows a relative reduction of about $77\%$ compared to the largest original REs, demonstrating both superior performance and a substantial reduction of the original CMF errors, as illustrated in Fig.~\ref{fig:sub_odd} for the odd data split. The right subfigure includes all REs from the $30$ different seeds, whereas the left subfigure shows the original REs for the same test points. About $95\%$ of all REs are below $13\%$ in predicting the overall mass flow rate, and the majority are below $3\%$. 
    Results on the even splits are highly similar.
    	
    Fig.~\ref{fig:sub_odd} shows that REs increase with GVF, consistent with prior work. To quantify it, Table~\ref{tab:gvf_bins} reports the same RE metrics as Table~\ref{tab:relative_errors} but separated into six GVF ranges for the best 4s and 60s models over the $30$ seeds. As GVF increases, the error distribution worsens: the 95th percentile rises and the proportions within $\le 10\%$ and $\le 5\%$ decline, with a few exceptions at the highest GVFs. The 4s CNN generally outperforms the 60s CNN---lower 95th percentiles in 9/12 ranges, lower medians in 10/12, and higher $\le 10\%$ coverage in 10/12---while the 60s CNN leads only in a few high-GVF ranges. The 4s split contains about $8$ to $14$ times more samples and higher unprocessed errors, thus its empirical 99th percentile and maximum can be larger simply from greater exposure to rare extremes: we emphasize the median, the 95th percentile, and coverage when assessing performance.
		
    \begin{table}[t!]
        \centering
        \setlength{\tabcolsep}{3pt}
        \caption{Relative error metrics by GVF-range, computed from all errors over 30 seeds (no per-seed averaging) for the 4s and 60s CNN models. Best results in bold; values in \%.}
        \begin{tabular}{c c c c c c>{\columncolor{yellow!20}} c c c c}
            \toprule
            \textbf{Split} & \textbf{Range} & \textbf{Model} & \textbf{n} & \textbf{Max} & \textbf{p99} & \textbf{p95} & \textbf{p50} & $\leq$ \textbf{10\%} & $\leq$ \textbf{5\%}  \\
            \midrule
            \multirow{14}{*}{Even} &
            \multirow{2}{*}{0--15~\%}  
                & 4s  & 13,140 & \textbf{27.2} & \textbf{12.8} & \textbf{8.35} & \textbf{2.40} & \textbf{97.2} & \textbf{82.5} \\
               & & 60s & 990 & 158 & 146 & 32.8 & 4.70 & 86.1 & 57.7 \\
            \addlinespace
            & \multirow{2}{*}{15--35~\%}  
                & 4s  & 9,450 & 25.0 & 18.7 & \textbf{13.6} & \textbf{3.21} & \textbf{91.4} & \textbf{69.3} \\
              &  & 60s & 1050 & \textbf{22.0} & \textbf{16.3} & 14.0 & 5.28 & 83.9 & 47.0 \\
            \addlinespace
            & \multirow{2}{*}{35--50~\%}  
                & 4s  & 2,850 & 29.8 & 25.8 & 22.5 & \textbf{3.74} & \textbf{83.6} & \textbf{63.4} \\
              &  & 60s & 300 & \textbf{21.0} & \textbf{18.6} & \textbf{15.2} & 5.88 & 79.3 & 42.3 \\
            \addlinespace
            & \multirow{2}{*}{50--70~\%}  
                & 4s  & 5,580 & 31.0 & \textbf{17.7} & \textbf{12.8} & \textbf{4.28} & \textbf{88.1} & \textbf{56.3} \\
              &  & 60s & 420 & \textbf{25.6} & 21.6 & 16.4 & 7.13 & 68.8 & 36.7 \\
            \addlinespace
            & \multirow{2}{*}{70--80~\%}  
                & 4s  & 1,230 & 44.4 & 34.1 & 25.5 & 8.41 & 58.5 & 23.9 \\
              &  & 60s & 120 & \textbf{24.8} & \textbf{22.8} & \textbf{16.5} & \textbf{5.81} & \textbf{74.2} & \textbf{40.8} \\
            \addlinespace
            & \multirow{2}{*}{80--95~\%}  
                & 4s  & 1,230 & 43.4 & 31.9 & \textbf{18.9} & \textbf{4.63} & \textbf{79.3} & \textbf{53.6} \\
              &  & 60s & 180 & \textbf{36.8} & \textbf{31.7} & 23.9 & 7.63 & 61.7 & 36.1 \\
            
            \midrule
            
            \multirow{14}{*}{Odd} &
            \multirow{2}{*}{0--15~\%}  
                & 4s  & 14,820 & \textbf{24.7} & \textbf{12.9} & \textbf{7.91} & \textbf{2.24} & \textbf{97.6} & \textbf{82.6} \\
                & & 60s &  930 & 25.8 & 21.6 & 14.6 & 4.40 & 87.9 & 57.4 \\
            \addlinespace
            & \multirow{2}{*}{15--35~\%}  
                & 4s  & 9,030 & \textbf{23.0} & \textbf{16.4} & \textbf{13.3} & \textbf{3.02} & \textbf{89.2} & \textbf{70.6} \\
                & & 60s &  840 & 42.5 & 23.4 & 17.5 & 4.96 & 74.5 & 50.4 \\
            \addlinespace
            & \multirow{2}{*}{35--50~\%}  
                & 4s  & 3,900 & \textbf{20.3} & \textbf{14.0} & \textbf{11.3} & \textbf{3.80} & \textbf{92.1} & \textbf{63.4} \\
               &  & 60s &  360 & 40.0 & 32.8 & 21.9 & 5.71 & 68.1 & 45.8 \\
            \addlinespace
           &  \multirow{2}{*}{50--70~\%}  
                & 4s  & 3,270 & \textbf{30.9} & \textbf{17.0} & \textbf{13.5} & \textbf{3.11} & \textbf{88.8} & \textbf{68.7} \\
              &   & 60s &  450 & 44.5 & 33.6 & 18.4 & 8.10 & 61.1 & 31.3 \\
            \addlinespace
            & \multirow{2}{*}{70--80~\%}  
                & 4s  & 1,020 & 27.3 & 21.9 & \textbf{15.7} & \textbf{7.7} & \textbf{69.8} & 25.0 \\
              &   & 60s &  150 & \textbf{22.3} & \textbf{20.4} & 17.4 & 9.4 & 52.0 & \textbf{30.0} \\
            \addlinespace
            & \multirow{2}{*}{80--95~\%}  
                & 4s  & 1,980 & 55.8 & 43.9 & 37.8 & 6.34 & 61.1 & 43.9 \\
              &   & 60s &  120 & \textbf{19.5} & \textbf{17.2} & \textbf{15.0} & \textbf{4.15} & \textbf{82.5} & \textbf{55.0} \\
              \bottomrule
              \multicolumn{10}{p{251pt}}{Here $n$ is the total number of samples (test points multiplied by 30 seeds).}
        \end{tabular}
        \label{tab:gvf_bins}
    \end{table}

    \subsubsection{Complementary Metrics}
    All evaluation metrics other than REs are presented in Table~\ref{tab:eval_metrics} for the 4s and 60s CNNs, along with the original metrics computed using the apparent CMF outputs as the predicted $\hat{y}$. Both NRMSE values and the $R^2$ indicate a very strong model performance when viewed independently, although they may be somewhat misleading as the original counterparts already showed high values, for instance an $R^2$ of $0.94$. These metrics alone are therefore not ideal for capturing the true difficulty of the problem, but they do suggest that the model itself is sound and indicate a small improvement.
    Comparing the MAE and MedAE shows that while there are some large errors, the median error is significantly lower and may therefore be a more accurate representation of the general model performance, especially if a few outliers are acceptable. The percentile metrics MAPE and MedAPE are both greatly reduced, particularly for the 4s CNN, suggesting that this model is more accurate, robust, and consistent across a variety of conditions.
    Overall, Table~\ref{tab:eval_metrics} shows that several complementary metrics are needed to fully characterize model performance and robustness, and that using short-time averages is beneficial.

    \begin{table}[t!]
		\caption{Average metrics for the best 4s and 60s models (CNN) across seeds $\pm$ stand. deviations. Best results in bold.}
		\setlength{\tabcolsep}{3pt}
		\centering
		\begin{tabular}{ll rl rl >{\centering\arraybackslash}p{0.8cm}>{\centering\arraybackslash}p{0.6cm}}
			\toprule
			\multirow{2}{*}{\textbf{Split}}&  \multirow{2}{*}{\textbf{Metric}}& \multicolumn{4}{c}{\textbf{CNN}} &  \multicolumn{2}{c}{\textbf{Original data}} \\
			\cmidrule(lr){3-6} \cmidrule(lr){7-8}
			& & \multicolumn{2}{c}{\textbf{4s}} & \multicolumn{2}{c}{\textbf{60s}} &  \textbf{4s} & \textbf{60s} \\
			\midrule
			\multirow{6}{*}{Even}
			& rNRMSE & \textbf{0.03} & $\pm$ \textbf{0.002} & 0.05 & $\pm$ 0.004 & 0.08 & 0.08  \\
			& avgNRMSE  &  \textbf{0.05} & $\pm$ \textbf{0.004} & 0.09 & $\pm$ 0.008  & 0.14 & 0.14 \\
			& MAE [\si{\kilogram\per\hour}] &  \textbf{265} & $\pm$ \textbf{20.7} & 440 & $\pm$ 45.06 & 700 & 736 \\
			& MedAE [\si{\kilogram\per\hour}] &  \textbf{183} & $\pm$ \textbf{21.6} & 304 & $\pm$ 46.28 & 462 & 506 \\
			& MAPE [\%] & \textbf{4.37} & $\pm$ \textbf{0.58} & 7.70 & $\pm$ 0.84 &15.5 & 16.4 \\
			& MedAPE [\%] & \textbf{3.14} & $\pm$ \textbf{0.41} & 5.37 & $\pm$ 0.99 & 8.61 & 9.50   \\
			& $R^2$ &  \textbf{0.99} & $\pm$ \textbf{0.001} & 0.98 & $\pm$ 0.004 & 0.94 & 0.94 \\
			\midrule
			\multirow{6}{*}{Odd}
			& rNRMSE &  \textbf{0.03} & $\pm$ \textbf{0.002} & 0.04 & $\pm$ 0.007 & 0.07 & 0.07 \\
			& avgNRMSE  &  \textbf{0.05} & $\pm$ \textbf{0.003} & 0.09 & $\pm$ 0.014 & 0.12 & 0.14 \\
			& MAE [\si{\kilogram\per\hour}]&  \textbf{246} & $\pm$ \textbf{22.5} & 409 & $\pm$ 86.05 & 599 & 674 \\
			& MedAE [\si{\kilogram\per\hour}]&  \textbf{179} & $\pm$ \textbf{28.1} & 317 & $\pm$ 95.7 & 397 & 409 \\
			& MAPE [\%] & \textbf{ 4.25} & $\pm$ \textbf{0.58} & 7.00 & $\pm$ 1.58  & 14.6 & 14.7 \\
			& MedAPE [\%] & \textbf{2.90} & $\pm$ \textbf{0.45} & 5.55 & $\pm$ 1.42 & 7.00   &  10.0 \\
			& $R^2$ &  \textbf{0.99} & $\pm$ \textbf{0.001} & 0.98 & $\pm$ 0.007  &0.96 & 0.96 \\
			\bottomrule
		\end{tabular}
		\label{tab:eval_metrics}
    \end{table}
	
    \vspace*{2\baselineskip}
    \subsection{Robustness Analysis}
    Models trained on the odd split generally perform slightly better than those on the even split, possibly because of the larger number of samples in the $0$--$15\%$ GVF range. Nonetheless, the results are similar, as shown in \Cref{tab:relative_errors,tab:gvf_bins,tab:eval_metrics}. 
    All previous comments regarding the benefits of sequence models and short-time averages hold for both splits. Despite the group-parity variability, especially evident in Table~\ref{tab:gvf_bins}, the overall results are consistent across splits, indicating that the findings are not due to a favorable split.
    	
\begin{figure}[t!]
    \centerline{\includegraphics[width=\columnwidth]{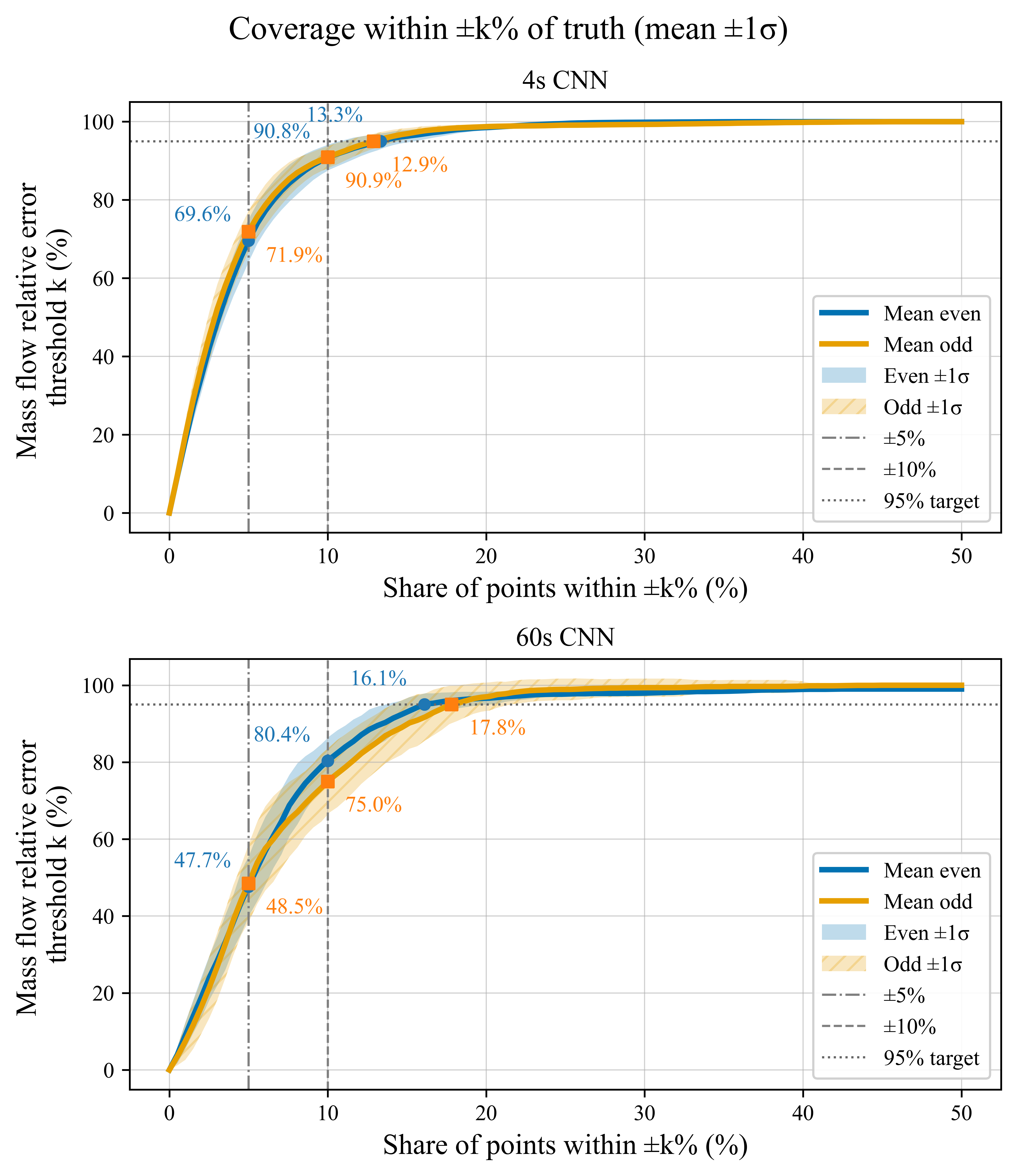}}
    \caption{Coverage percentiles averaged per seed for both data splits. Results are shown for the 4s CNN (top) and 60s CNN (bottom) models. Shading indicates $\pm 1$  standard deviation across seeds.}
    \label{fig:REaveraged_per_seed}
    \end{figure}

    Fig.~\ref{fig:REaveraged_per_seed} shows the average percentage of REs, across all seeds, that are within a certain percentile threshold for the 4s and 60s CNN models. The 4s CNN has only small percentage-point differences between the splits, while the between-seed variability is minimal and nearly identical across splits, as indicated by the overlapping shaded areas in the top subfigure. The 60s CNN exhibits greater between-seed and group-parity variability, suggesting that it is not as robust as the 4s CNN. Furthermore, the standard deviations in Table~\ref{tab:eval_metrics} support the consistency of the results by indicating a limited variability in the evaluation metrics across seeds for the 4s CNN.
    Together, these results confirm that the 4s CNN provides consistently robust performance across both data splits and random seeds.
		
	\subsection{Model Complexity}
	Table~\ref{tab:model_complexity} shows different metrics used to assess the complexity of all three models when trained on the 4s downsampled data. Latency is often the most important metric in real-time applications as they require instant responses, and the MLPw can thus be a solid alternative to the CNN if latency is prioritized over a small loss in accuracy. The MLPw was consistently the fastest of the three models for all downsampling intervals, which is notable given that its network structure is larger than that of the MLP. Nevertheless, the two MLPs only differ by $3~\si{\micro\second}$, and the CNN still has a latency of $170~\si{\micro\second}$, so they are all relatively fast, as well as simple compared to many modern ML models.
	
	\begin{table}[t!]
		\caption{Complexity metrics for models trained on the 4~\si{\second} data. Latency refers to the mean across seeds $\pm$ stand. deviation.}
		\setlength{\tabcolsep}{3pt}
		\centering
		\begin{tabular}{lcc rl rl}
            \toprule
            \multirow{2}{*}{\textbf{Model}} & \multirow{2}{*}{\textbf{Parameters}} & \multirow{2}{*}{\shortstack{\textbf{MACs per} \\ \textbf{example}}} & 
            \multicolumn{4}{c}{\textbf{Latency [\si{\micro\second} per example]}} \\
            \cmidrule(lr){4-7}
            & & &  \multicolumn{2}{c}{Even} & \multicolumn{2}{c}{Odd} \\
            \midrule
            MLP       &   57  &   96    &   
            30  &$\pm$ 1.3 & 31  &$\pm$ 1.3 \\
            MLPw      &  433  &  832    &   
            27  &$\pm$ 2.6 & 28  &$\pm$ 3.1 \\
            CNN       &  809  &  6,528   &  
            170 & $\pm$ 46  & 160 & $\pm$ 39  \\
            \bottomrule
        \end{tabular}
		\label{tab:model_complexity}
	\end{table}
	
\section{Conclusion and Future Work}\label{sec:conclusion}
	This study demonstrated that correcting CMF measurements by exploiting temporal dependencies and applying short-time averaging within experiments yields significantly improved results in terms of lower REs, MAPEs, NRMSEs, and $R^2$ values. Three models (an MLP, a MLPw, and a 1D-CNN) were compared across various downsampling intervals. The best performance was obtained by the CNN sampled at $0.25~\si{\hertz}$, achieving about $95\%$ of REs below $13\%$, an rNRMSE of $0.03$, and a MAPE of about $4.3\%$. The models were trained on three-phase air--water--oil flow data covering a wide range of experimental conditions, and robustness was ensured by training and evaluating on two data splits and 30 random seeds. Further research may explore different network structures and models, as well as time-aware feature engineering to improve the accuracy, especially for higher GVFs. 

	\bibliography{references}

	\begin{IEEEbiography}[{\includegraphics[width=1in,height=1.25in,clip,keepaspectratio]{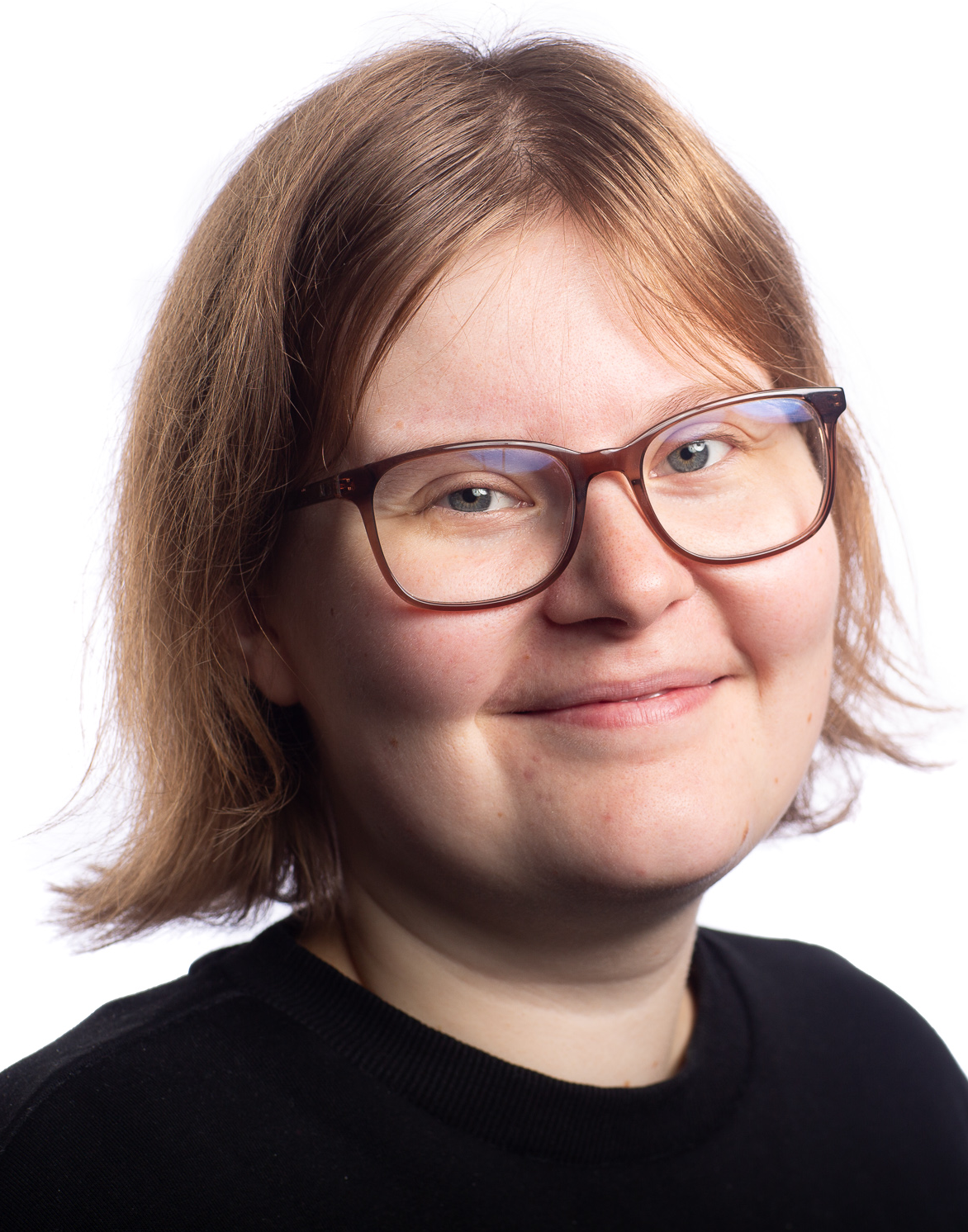}}]{Amanda Nyholm} received the B.Sc. degree in mathematics and M.Sc. in mathematical statistics from Lund University, Sweden, in 2020 and 2022, respectively. She is currently pursuing the Ph.D. with the Department of Electronic Systems, Norwegian University of Science and Technology (NTNU), Trondheim, Norway, within the Signal Processing Research Group. 
    
    Her research interests include multivariate time series, signal processing, machine learning, and statistics.
	\end{IEEEbiography}
    
    \begin{IEEEbiography}[{\includegraphics[width=1in,height=1.25in,clip,keepaspectratio]{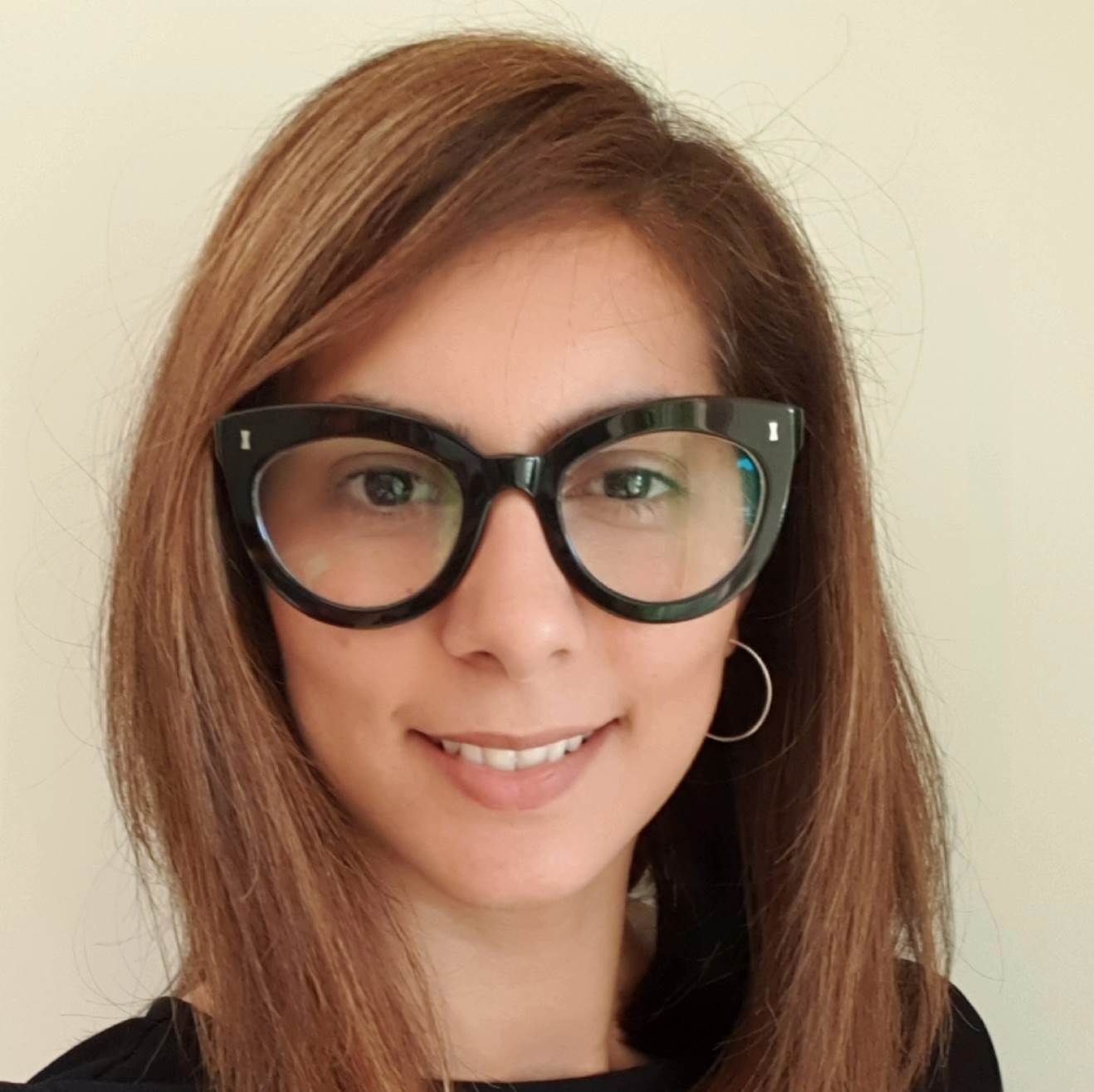}}]{Yessica Arellano} received the B.E. and
    	M.Sc. degrees from La Universidad del Zulia,
    	Venezuela, the second M.Sc. (Hons.) degree
    	from Robert Gordon University, U.K., and the
    	Ph.D. degree from Coventry University, U.K.,
    	in 2020, focused on advanced multiphase
    	flow monitoring via electromagnetic
    	measurements. 
    	
    	She is currently a Senior Research Scientist with the Department of Gas Technology, SINTEF Energy Research, Norway, specializing in flow measurement technologies for carbon dioxide transport systems. Yessica is also the inventor behind two patented innovations for apparatuses and method for measuring multiphase flow.
	\end{IEEEbiography}

    \begin{IEEEbiography}[{\includegraphics[width=1in,height=1.25in,clip,keepaspectratio]{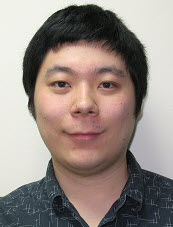}}]{Jinyu Liu} received the B.Eng. degrees from Tianjin University, China, and the University of Kent, Canterbury, U.K., in 2012, the M.Sc. degree from the University of Southampton, U.K., in 2013, and the Ph.D. degree in Electronic Engineering from the University of Kent, Canterbury, U.K., in 2019.
    
    He is currently a Research Engineer in the R\&D Department, KROHNE Ltd., Wellingborough, U.K. His research interests include Coriolis flowmeter technology, multiphase flow measurement, condition monitoring of industrial processes, and digital signal processing.
	\end{IEEEbiography}

    \begin{IEEEbiography}[{\includegraphics[width=1in,height=1.25in,clip,keepaspectratio]{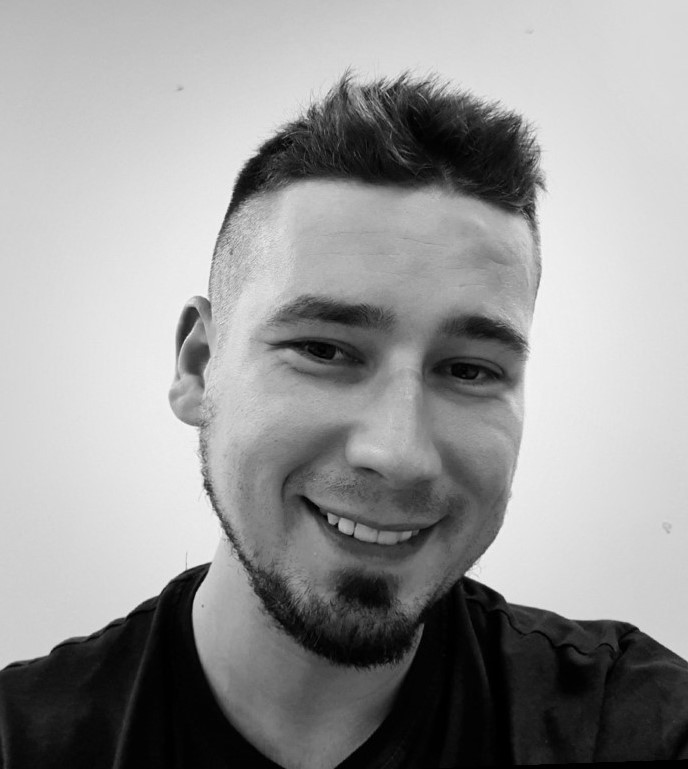}}]{Damian Krakowiak} received the B.Sc. (Hons.) in Electrical \& Electronic Engineering from Glasgow Caledonian University, U.K., in 2012.
    
    He is currently Senior Metrology Manager and Custody Transfer Approvals Engineer at KROHNE Ltd., Wellingborough, U.K., specialising in metrology, fiscal measurement, and ISO 17025 compliance. With prior technical leadership at TÜV SÜD, he has extensive experience in flow measurement, calibration automation, and quality system development. His work focuses on advancing metrological standards and ensuring precision within complex fiscal and custody transfer applications. 
	\end{IEEEbiography}

    \begin{IEEEbiography}
    [{\includegraphics[width=1in,height=1.25in,clip,keepaspectratio] {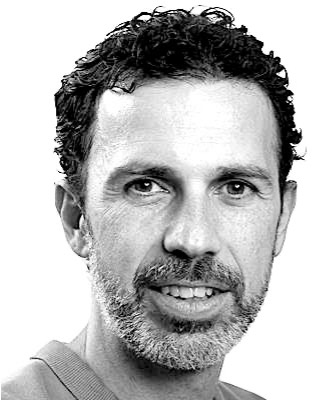}}]
    {Pierluigi~Salvo~Rossi} (SM'11) was born in Naples, Italy, in 1977. 
    He received the Dr.Eng. degree (\emph{summa cum laude}) in telecommunications engineering and the Ph.D. degree in computer engineering from the University of Naples ``Federico II,'' Italy, in 2002 and 2005. 
    
    He is currently a Full Professor and the Deputy Head with the Department of Electronic Systems, Norwegian University of Science and Technology (NTNU), Trondheim, Norway. 
    He is also a part-time Senior Research Scientist with the Department of Gas Technology, SINTEF Energy Research, Norway. 
    He worked with the University of Naples ``Federico II,'' Italy, with the Second University of Naples, Italy, with NTNU, Norway, and with Kongsberg Digital AS, Norway. 
    He held visiting appointments with Drexel University, USA; Lund University, Sweden; NTNU, Norway; and Uppsala University, Sweden. 
    His research interests fall within the areas of communication theory, data fusion, machine learning, and signal processing. 
    
    Prof. Salvo Rossi was awarded Exemplary Senior Editor of the \textsc{IEEE Communications Letters} in 2018.
    He is (or has been) on the Editorial Board of the \textsc{IEEE Sensors Journal},
    the \textsc{IEEE Open Journal of the Communications Society}, 
    the \textsc{IEEE Transactions on Signal and Information Processing over Networks}, 
    the \textsc{IEEE Communications Letters} and the \textsc{IEEE Transactions on Wireless Communications}.
    \end{IEEEbiography}

\vfill
	
\end{document}